\documentstyle[preprint,aps]{revtex}

\begin{document}
\draft
\title{Enhanced Electron Pairing in a Lattice of Berry Phase Molecules}
\author{
Nicola Manini\thanks{Email: manini@sissa.it}}
\address{
International School for Advanced Studies (SISSA),
via Beirut 2-4, I-34013 Trieste, Italy}
\author{
Erio Tosatti\thanks{Email: tosatti@sissa.it}}
\address{
International School for Advanced Studies (SISSA), via Beirut 2-4, I-34013
Trieste, Italy\\ and International Centre for Theoretical Physics (ICTP),
P.O. BOX 586, I-34014 Trieste, Italy}
\author{
Sebastian Doniach\thanks{Email: doniach@drizzle.stanford.edu}}
\address{Dept. Applied Physics, Stanford University, Stanford CA 94305 USA}
\date{July 07, 1994}
\maketitle

\begin{abstract}
We show that electron hopping in a lattice of molecules possessing a Berry
phase naturally leads to pairing. Our building block is a simple molecular
site model inspired by C$_{60}$, but realized in closer similarity with
Na$_3$. In the resulting model electron hopping must be accompanied by
orbital operators, whose function is to switch on and off the Berry phase
as the electron number changes. The effective hamiltonians (electron-rotor
and electron-pseudospin) obtained in this way are then shown to exhibit a
strong pairing phenomenon, by means of 1D linear chain case studies. This
emerges naturally from numerical studies of small $N$-site rings, as well
as from a BCS-like mean-field theory formulation. The pairing may be
explained as resulting from the exchange of singlet pairs of orbital
excitations, and is intimately connected with the extra degeneracy implied
by the Berry phase when the electron number is odd. The relevance of this
model to fullerides, to other molecular superconductors, as well as to
present and future experiments, is discussed.
\end{abstract}

% Strongly correlated el., SC Theory , Fullerenes

\section{Introduction}
\label{intro}

A significant feature of the physics of unconventional superconductors such
as the cuprates is the constraint imposed on the motion of the charge
carriers by the background degrees of freedom, i.e. the spins in the case
of the cuprate superconductors.

In this paper we focus attention on a new class of constraints imposed on
the motion of conduction electrons by the Berry phase\cite{berry}, or
molecular Aharonov-Bohm phase \cite{mead}, which can arise in molecular
crystals with large on--site degeneracies.  In general, for this to be the
case the symmetries of electron and vibron states must match appropriately
and, moreover, the number of electrons needs to be odd.

As an explicit example, we have demonstrated elsewhere \cite{assa1} the
presence of a Berry phase in negatively charged fullerene ions C$_{60}^-$ ,
C$_{60}^{3-}$, C$_{60}^{5-}$ ensuing from a Dynamic Jahn--Teller (DJT)
effect arising from coupling between the partly occupied $t_{1u}$ orbital
and the $H_g$ vibron modes.

In such a situation a physical electron (or hole) may be regarded as a
composite particle, made up of the bare electron plus the geometrical phase
which accompanies it when sitting, unpaired, (or more generally in a state
with an odd number of electrons) on a given molecule. By contrast, either
the absence of electrons or presence of a pair of electrons will eliminate
the Berry phase on that molecule. We argue below that the kinematical
constraints imposed by the Berry phase can be a factor capable of tilting
the balance in favor of pairing, even in the presence of repulsive
interactions \cite{tosatti}. The way this works is for a pair of electrons
on one molecule to gain energy by each tunnelling off onto neighbors with
accompanying vibron excitations before coming together again. A single
electron on the other hand will tunnel accompanied by its vibron
excitations, so the electrons will gain less energy by tunnelling as
individuals than will be gained by tunnelling as pairs.

In this paper, we propose a class of simple coupled electron--rotor models
which we believe capture the essential physics introduced by the Berry
phase constraint. Omitting at this stage the complications of real
C$_{60}^{n-}$ anions, our model lattice hamiltonian is instead directly
inspired by the simpler and well known \cite{delac} strong--coupling Berry
phase molecule Na$_3$.

In general there are an infinite number of rotor states on each molecular
site. To simplify further, we shall truncate to a 3--state model for the
rotors.  The resulting many--site Hamiltonian will be shown to take the
form
\begin{equation}
	H=-\frac{t}{2} \sum_{<i,j>,\sigma} 		 c^\dagger_{i
,\sigma} c_{j,\sigma} 		(S^+_i S^-_j + h.c.)
\label{spin1}
\end{equation}
where $S^+,S^-$ are raising and lowering operators for a spin--1 vibron
manifold for each molecule, and $<i,j>$ denote first neighbors.

To test the pairing properties of (\ref{spin1}), we include a Hubbard
repulsion term
\begin{equation}
H_U=U\sum_{i}n_{i\uparrow}n_{i\downarrow}.
\end{equation}
We then study the half--filled state of the model numerically for small
systems (4 to 8 site rings), and also by a mean field BCS--type calculation
for the 1--D chain.

The final conclusion is that the Berry phase coupling is found to be
greatly beneficial to electron pairing, at least within the simple 1D
lattice studied so far. Pairing, in particular, appears to prevail and to
survive even in presence of a repulsive Hubbard $U$, up to values
$U\approx t$.

In Section \ref{model} we will introduce the main concepts and build our
working hamiltonian, representing an idealized lattice of Berry phase
molecules.

Section \ref{numeric} is devoted to a numerical   study of the 4,
6 and 8--site 1D chain clusters, where correlations can be studied exactly,
and the presence of pairing is demonstrated.

Section \ref{meanf} discusses a mean field BCS--type formulation for the
infinite 1-D chain.

We close with a discussion section, where a number of interesting open
issues are also briefly presented.

\section{The Model}
\label{model}

Our model system is a regular lattice of molecules. Each molecule possesses
initially a degenerate orbital, an odd number of electrons, and a dynamical
Jahn-Teller (DJT) coupling (with Berry phase) to a local vibration, also
degenerate, of pseudorotational type. For simplicity, we stick to the case
of one electron in a doubly degenerate orbital and one rotor, which
provides the simplest case of Berry phase coupling. The electron can hop
from a molecule to the next one in the usual fashion, conserving ordinary
spin. We will generally also include an on-site (``intra-molecular'')
electron-electron repulsion $U$, so that, for a nondegenerate level and no
rotor coupling, we would have just an ordinary Hubbard model. Berry phase
coupling to the rotors is the new ingredient giving rise to peculiar
selection rules for electron hopping between two molecules in different
rotor states. In the following, we describe successively the on-site
hamiltonian, and the full lattice hamiltonian inclusive of electron
hopping.

\subsection{On-site hamiltonian: modeling a Berry phase molecule}
\label{1site}

We consider at each site a molecule with a partly occupied doubly
degenerate electronic state. Suppose this orbital interacts via linear DJT
coupling with a doubly degenerate vibration. A practical example (not
relevant for superconductivity) of such a situation is the Na$_3$ molecule.
A single unpaired electron occupies the doubly degenerate electronic
molecular orbital $E \equiv (E_x,E_y)$, and the doubly degenerate vibron is
a pseudorotation of the Na$_3$ triangular structure. Direct spectroscopical
evidence has been found \cite{delac}, showing that the formalism we present
here (supplemented by quadratic couplings which are omitted here for
simplicity) correctly describes the dynamics of this system.

The single-molecule, one-electron, linear coupling case, is a classic
Jahn-Teller textbook problem \cite{englman}. When only one electron occupies
the degenerate state, the lowest order electron-vibron hamiltonian
can be written:
\begin{equation}
	H =  {{\hbar \omega} \over 2}
           \left[{{\partial^2} \over {\partial q^2_1}} +
	 {{\partial^2} \over {\partial q^2_2}} + q_1^2 + q_2^2 \right]
	\pmatrix{ 1 & 0 \cr 0 & 1 }
	+ g {{\hbar \omega} \over 2} \pmatrix{ -q_1 & q_2 \cr q_2 & q_1 }
\label{ham1}
\end{equation}
where the $2\times2$ matrices span the twofold $E$ electronic level, and
$q_i$ are the vibrational normal coordinates in the vibron space; the
second term is the coupling between vibron and electron, of dimensionless
strength $g$.

The problem is rotationally invariant in the ($q_1$,$q_2$) space. It is
therefore conveniently rewritten by introducing polar coordinates $ q ,
\phi $ in the vibron space and a pseudospin $\frac{1}{2}$ representation in
the electron space:
\begin{equation}
\left| {\mp \left. \frac{1}{2} \right\rangle} \right.=\pm {i \over {\sqrt
2}}\left(
{\left| {E_x \left. {} \right\rangle} \right.\mp i\left| {E_y
\left. {} \right\rangle} \right.} \right)
\label{baschange}
\end{equation}

In terms of these quantities the hamiltonian is expressed as:
\begin{equation}
H = {{\hbar \omega } \over 2}\left[ {-{{\partial ^2} \over {\partial
q^2}}-{1 \over q}{\partial  \over {\partial q}}-{1 \over {q^2}}{{\partial ^2}
\over {\partial \phi ^2}}+q^2} \right]+{{g\hbar \omega } \over 2}\left(
{\matrix{0&{qe^{+i\phi }}\cr
{qe^{-i\phi }}&0\cr
}} \right)
\label{ham2}
\end{equation}

A new total angular momentum $j$, defined as
\begin{equation}
j=\hbar ^{-1}\left( {q_1p_2-q_2p_1} \right)-{1 \over 2}\sigma _z=-i{\partial
\over {\partial \phi }}-\left( {\matrix{\frac{1}{2}&0\cr
0&{-\frac{1}{2}}\cr
}} \right) \ ,
\label{jdef}
\end{equation}
is found to commute with $H$ \cite{englman}. Note that $\sigma _z$ is only
a {\em pseudospin} $\frac{1}{2}$ spanning in reality the twofold {\em
orbital} state, and should not be confused with the {\em true} spin, which
is ignored at this stage. Because of this pseudospin $\frac{1}{2}$ term,
the eigenvalues of $j$ are half-odd integer, an amusing anomaly first
pointed out by Herzberg and Longuet-Higgins \cite{lh} for the case of the
triangular molecule. This fractionalization can be seen as a manifestation
of a Berry phase \cite{berry} of $\pi$, which the vibrons pick up from the
electron degeneracy \cite{delac}.

Diagonalization of (\ref{ham2}) must in general be done numerically
\cite{englman,uehara}. In the limit of strong coupling ($g>>1$), however, the
massive radial $q$-motion can be approximately separated from the
$\phi$-pseudorotation quantized by $j$, and both can be solved analytically in
the form of an oscillator and a free rotor respectively.

The resulting spectrum is classified according to $j$ and $\nu$, the quantum
number coming from the quantization of the radial massive motion motion:
\begin{equation}
	E(v,j) = \hbar \omega (\nu  + {1\over 2}) +
		{{2 \hbar \omega }\over {g^2}}\left(
		 j^2 + {1\over 4}
		\pm \sqrt{j^2 + {{g^8}\over{64}}} ~ \right) \ , \
		\nu = 0,1,2,... \ , \ j=\pm \frac{1}{2}, \pm \frac{3}{2}, ...
\label{2.1}
\end{equation}
Since we shall be concerned only with the low lying rotor states, we can
forget the massive boson ladder. Furthermore, we will express all energies in
units of the pseudorotational quantum $\Omega:={{2 \hbar \omega }\over {g^2}}$.
% NOTICE THAT while omega scales with m^{-1/2}, Omega scales with m^{-1}!!!

The corresponding wavefunctions in the strong coupling limit are:
\begin{equation}
\psi_{\nu,j} (q,\phi) = \psi_\nu (q) \left( {\matrix{{\cos \theta_j
e^{i(j+{1\over2})\phi}}\cr
{\sin \theta_j e^{i(j-{1\over2})\phi}}\cr}} \right)
\ j \mbox{ half-odd integer}
\label{psihc}
\end{equation}
where $\psi_\nu (q)$ is the appropriate harmonic oscillator wavefunction, and
$\theta_j$ is a pseudospin mixing angle
\begin{equation}
	\tan \theta_j=8{{-j \pm \sqrt{j^2 + {{g^8}\over{64}}} } \over {g^4}}
\label{thetadef}
\end{equation}

In the $g \rightarrow \infty$ (strong coupling) limit $\theta_j$  tends to $\pm
{\pi \over 4}$. The energy can be expanded as
\begin{equation}
	E(j)= \pm {{g^4}\over 8} + j^2 + {1\over 4} \pm 4{j^2\over{g^4}}+\cdots
\label{2.2}
\end{equation}

At low energies, we consider only the rotor states
\begin{equation}
\psi_{j} (\phi) = 2^{-\frac{1}{2}} \left( {\matrix{
{e^{i(j+{1\over2})\phi}} \cr {-e^{i(j-{1\over2})\phi}}\cr}} \right)
	\ , \ j = \pm \frac{1}{2},\pm \frac{3}{2}, \pm \frac{5}{2},...
\label{psih}
\end{equation}
with energy:
\begin{equation}
	E(j) =   j^2  \ ,
	\ \ j = \pm \frac{1}{2},\pm \frac{3}{2}, \pm \frac{5}{2},... \ ,
\label{enh}
\end{equation}
where we have omitted as usual the $ -\frac{g^4}{8} $ offset
contribution (polaron energy shift), but also the extra zero point energy
1/4 required by the Berry phase \cite{assa1}. This is our simplified model
for the one-electron Berry molecule.

To study electron hopping among different molecules, we also need an
equivalent description for molecular occupancy different from one. When two
electrons occupy the molecular orbital in a singlet state, the DJT distortion
is still present. However, the orbital phases for the two electrons cancel each
other and a Berry phase is no longer present.

For uniformity with the one-electron case, we will still label for $n$=2 the
pseudorotational levels with $j$. Cancellation of the geometrical phase now
requires  $j$ to be integer. The assumed spectrum is simply that of a standard
free pseudorotor:
\begin{equation}
E(j) = j^2 \ , \ j = 0,\pm 1,\pm 2, ...
\label{eni}
\end{equation}
with wavefunctions
\begin{equation}
\psi_j (\phi) = e^{i j \phi} \ , \ j = 0,\pm 1,\pm 2, ...
\label{psii}
\end{equation}
In the energy eigenvalue (\ref{eni}) we are again omitting the polaron
energy gain $-{{g^4}\over 2}$, (units of $\Omega$).

In the $n$=0 case where no electron is present, no Jahn-Teller effect and
thus no pseudorotor either. However we really would like to mimic an
electron-hole symmetric situation, like for example going C$_{60}^{2-}$ or
C$_{60}^{4-}$ with respect to C$_{60}^{3-}$. For this reason we assume even
for zero electrons a pseudorotation (\ref{eni},\ref{psii}) identical to
that of the two electron case.

We finally discard occupancies higher than two; for instance, we may
suppose that, if the average occupancy is one, states with $n>2$ are strongly
suppressed by Coulomb repulsions.

To summarize, if $n$ is the number of electrons present in the degenerate
orbital, $2j$ assumes even or odd values according to whether $n$ is even
or odd. The $j$-dependence of the energy eigenvalues is quadratic, like in
a free (pseudo-)rotor.  The full quantum state of such a molecule, in the
limit considered, is described by a set of three quantum numbers which for
convenience we define as integers:
\begin{eqnarray}
	n 	& \mbox{, (occupancy)}		\nonumber\\
	m = 2 j & \mbox{, (rotor state)}	\nonumber\\
	\sigma = 2 m_s& \mbox{, (spin state)}
\label{2.5}
\end{eqnarray}
where $m_s = \pm \frac{1}{2}$ is the z component of true electron spin.
Their allowed values are constrained in the form:
\begin{eqnarray}
n 	& = & 0, 1, 2			\nonumber\\
m 	& = & 2 l + \sin (\pi n) \ , \ l=0,\pm1,\pm2,\pm3 \cdots \nonumber\\
\sigma	& = & \pm \sin (\pi n)
\label{const}
\end{eqnarray}

With reference to the physics of C$_{60}^{n-}$, discussed in Ref.
\cite{assa1}, the similarity with the present model should be clear. In
that case, in particular, the ground state has even $L$ for even $n$, and
odd $L$ for odd $n$.

\subsection{Inter-site hamiltonian $\rightarrow$ hopping between Berry
molecules.}

To allow electrons to move among sites, we need to specify how the hopping
process is affected by the $j$ quantum numbers on each site. Therefore we
begin considering one electron hopping between two neighbouring molecules.
With sufficiently high point symmetry, such as one has for the linear
chain, we will have hopping between $\left| {E_{x,1}} \right\rangle$ and
$\left| {E_{x,2}} \right\rangle$ and between $\left| {E_{y,1}}
\right\rangle$ and $\left| {E_{y,2}} \right\rangle$ only, with amplitude
$t_x$ and $t_y$ respectively \cite{note1}.  If for instance the twofold
degenerate state is associated to a p-orbital, then $t_x=t_{pp\sigma}$,
$t_y=t_{pp\pi}$, in Slater-Koster's notation\cite{staterkoster}. The hopping
hamiltonian then is
\begin{equation}
H_{kin}=\sum_{\sigma}
\left[t_x(c_{x,\sigma,1}^{\dagger}c_{x,\sigma,2}+h.c.)
+t_y(c_{y,\sigma,1}^{\dagger}c_{y,\sigma,2}+h.c.)\right]
\label{hhop}
\end{equation}

To characterize its behaviour, suppose we have on site 2 a spin up
electron, with $m_2$ (odd) molecular pseudospin, and we want
the hopping amplitude of this electron to site 1 with final spin up, and
pseudospin $m_1$ (also odd). We need to invert relation (\ref{baschange})
to express the fermionic operators in the pseudospin basis:

\begin{eqnarray}
c_{x,s}^{\dagger}={{c_{+,s}^{\dagger}-c_{-,s}^{\dagger}} \over {-i\sqrt2}}
\nonumber\\
c_{y,s}^{\dagger}={{c_{+,s}^{\dagger}+c_{-,s}^{\dagger}} \over {\sqrt2}}
\label{cxy}
\end{eqnarray}

The matrix element we compute depends also on the (even) pseudospin
of the empty sites $l_2$ (final state) and $l_1$ (initial state) in
the following way:
\begin{eqnarray}
	\left\langle{\matrix{1&0\cr m_1&l_2\cr \uparrow&0\cr}}
	\right| H_{kin}
	\left|{\matrix{0&1\cr l_1&m_2\cr 0&\uparrow\cr}}
	\right\rangle = &
	{{t_x+t_y} \over 2}\left( {{\delta _{m_1+1
	,l_1}\delta _{l_2,m_2+1}+\delta _{m_1-1,l_1}
	\delta_{l_2,m_2-1}} \over 2} \right)+ \nonumber\\
&	{{t_x-t_y} \over 2}\left( {{\delta_{m_1+1,l_1}
	\delta _{l_2,m_2-1}+\delta _{m_1-1,l_1}
	\delta _{l_2,m_2+1}} \over 2} \right)
\label{kinel}
\end{eqnarray}

The deltas here originate through trivial orthogonality of angular momentum
wavefunctions (\ref{psih},\ref{psii}), while the overall ${1 \over 2}$ factors
originate from the $\sqrt2$ factors in (\ref{cxy}) and in (\ref{psih}).

If we limit ourselves to the special case of intermolecular interaction
with $t_y=t_x$ ($t_y=-t_x$), then we have an additional conservation
$l_1+m_2=m_1+l_2$ ($l_1-m_2=m_1-l_2$) of the total pseudospin, as shown by
Eq. (\ref{kinel}). In this cases, the spectrum separates into different
independent manifolds. For $t_y=t_x$, they correspond to different values
of total pseudospin. Even if somewhat unrealistic (for a real p state,
$t_x$ is larger and positive, $t_y$ smaller and negative) this is a very
convenient choice and we shall adopt it in the following. We have made a
check, to be described in Sec. \ref{numeric}, releasing
this restriction, which have satisfied us that the physics is not
fundamentally different in the more general case $t_y \neq t_x$.
Accordingly, we define a single effective hopping
\begin{equation}
	t={{t_x+t_y}\over 2}
\label{tdef}
\end{equation}
which in the absence of direct electron-electron interactions is the only
independent parameter of our model, the rotor energy quantum $\Omega$ being
unity.

Similar considerations and selection rules to those discussed above for the
process $|0,1> \to |1,0>$ apply to the case involving doubly occupied
sites, namely $|0,2> \to |1,1>$, or $|1,2> \to |2,1>$.

Now we have all the ingredients to place these ``Berry molecules'' on a
lattice to see the effect of local rotor coupling on electron hopping.

\subsection{Lattice of Berry molecules: the working hamiltonian}

As a first attempt to study this model on a lattice we consider a linear
chain of $N$ sites, with $N_{el}$ electrons moving on them. The basis we
consider is labelled by the set of all the $n_i, m_i, \sigma_i $, for $ i =
1,\cdots, N$, so that an arbitrary state is expandable on states:
\begin{equation}
\left| {\matrix{n_1	&n_2	 &\cdots &n_{N}		\cr
		m_1	&m_2	 &\cdots &m_{N} 	\cr
		\sigma_1&\sigma_2&\cdots &\sigma_{N}	\cr }}
\right\rangle \ .
\label{basis}
\end{equation}
These basis states are obtained by ordered applications of $N_{el}$ local
fermionic creation operators $c^\dagger_{j,\sigma}$ on a vacuum state where
no electrons are present, and setting the $m_i$ rotational quantum numbers
to values allowed by the Berry constraint (\ref{const}).

Hopping of an electron between site $i$ and site $j$ is, in this space, a
composite operation, since it implies $n_i \to n_i -1$, $n_j \to n_j +1$,
but also $m_i \to m_i \pm 1$, $m_j \to m_j \mp 1$
(\ref{const},\ref{kinel}). Hence, we write a general hopping hamiltonian
in second quantized language as:
\begin{equation}
   H_{kin} =  -{t\over 2} \sum_{<i,j>,\sigma}
		 c^\dagger_{i,\sigma} c_{j,\sigma}
	 	( L^+_i L^-_j + L^-_i L^+_j)	\ .
\label{gen_model}
\end{equation}
where the action of the operator $L^\pm_j$ is to raise (lower) the pseudospin
$m_j$ (really an orbital angular momentum) by one unit:
\begin{equation}
        L^\pm_j
\left| {\matrix{n_1     &\cdots	&n_{j}	&\cdots		\cr
                m_1     &\cdots &m_{j}	&\cdots		\cr
                \sigma_1&\cdots &\sigma_{j}&\cdots	\cr }} \right\rangle
=
\left| {\matrix{n_1     &\cdots	&n_{j}	&\cdots		\cr
                m_1     &\cdots &m_{j}\pm 1&\cdots	\cr
                \sigma_1&\cdots &\sigma_{j}&\cdots	\cr }}\right\rangle \ .
\label{Lpm}
\end{equation}

To the hopping hamiltonian $ H_{kin}$, we add an on-site rotor hamiltonian
$H_{rot}$, as well as an additional on-site electron-electron Hubbard
interaction term $H_U$
\begin{eqnarray}
	H_{rot}&=&\sum_{j=1}^N \frac{1}{4} \left( L^z_j \right)^2 \nonumber\\
	H_U &=&   U \sum_{j=1}^{N} n_{j,\uparrow} n_{j,\downarrow}	\ ,
\label{hloc}
\end{eqnarray}
the rotor energy contribution is at site $j$ due in the rotor state $m_j$
being
\begin{equation}
H_{rot~j}=\frac{1}{4}m_j^2 \ ,
\end{equation}
as in Eq. (\ref{enh},\ref{eni}), and having introduced a third operator
\begin{equation}
        L^z_j
\left| {\matrix{n_1     &\cdots	&n_{j}	&\cdots		\cr
                m_1     &\cdots &m_{j}	&\cdots		\cr
                \sigma_1&\cdots &\sigma_{j}&\cdots	\cr }} \right\rangle
= m_j
\left| {\matrix{n_1     &\cdots	&n_{j}	&\cdots		\cr
                m_1     &\cdots &m_{j}	&\cdots	\cr
                \sigma_1&\cdots &\sigma_{j}&\cdots	\cr }}\right\rangle \ .
\label{Lz}
\end{equation}
The commutation relations for these operators are $[L^+,L^-]=0$,
$[L^z,L^\pm]=\pm L^\pm$.

We stress here that the kinetic term $H_{kin}$ alone in Eq.
(\ref{gen_model}) is the relevant part of the new hamiltonian we want to
study.

The new electron operators $c^\dagger_{i,\sigma}L^\pm_i$ are diffrent from
the original ones, $c^\dagger_{x,\sigma,i}$, $c^\dagger_{y,\sigma,i}$ of
Eq. (\ref{hhop}). In particular, we have now a single band, instead of the
original double band problem. However, with the exclusion of all site
occupancies higher than 2, all matrix elements are the same in the two
descriptions.

By construction, the hamiltonian (\ref{gen_model}) conserves the
constraints (\ref{const}) among the quantum numbers, and can therefore be
diagonalized in the Hilbert space of states defined in (\ref{basis}).
The matrix elements of the kinetic term $H_{kin}$ (off diagonal) and of $
H_{rot} + H_U$ (diagonal) on the basis (\ref{basis}) are trivial, once
periodic boundary conditions (PBC) are applied to indexes. In some cases we
shall however need antiperiodic boundary conditions (ABC), replacing $t$
with $-t$ in the kinetic term involving sites 1 and $N$.

The Hilbert space of the problem is infinite-dimensional even for finite
$N$, due to the $m_i$ rotor quantum numbers, which are boundless. In the
numerical computations, we shall truncate the basis (\ref{basis}) by
choosing a cutoff energy $E_{cut}$, including only states having some local
energy smaller than this $E_{cut}$.  Unfortunately, the choice $H_{rot}
\leq E_{cut}$ is unfair with respect to singly occupied states, having
larger energy (at least by 1/4) than the unoccupied and doubly occupied
lowest-$j$ ones. To achieve better convergence even at relatively small
$E_{cut}$, we subtract this contribution and retain those states
satisfying:
\begin{equation}
        H_{rot} + \sum_j \left( \frac {n_{j,\uparrow} n_{j,\downarrow}}{2} -
                \frac {n_{j,\uparrow} + n_{j,\downarrow}}{4} \right)
		\leq E_{cut} \ ,
\label{2.9}
\end{equation}

The special case $E_{cut} =0$ is of very strong interest. Physically, this
corresponds to the limiting case $ t << \Omega$, where the intramolecular
rotor energy is much larger than the hopping energy.  In this limit, the
only allowed values for $m_j$ are 0 (even occupancy) and $\pm 1$ (odd
occupancy). The resulting model has six states per site, two corresponding
to even (0 and 2) occupancies, and four to the $2\times 2$ combinations of
spin $\sigma_j$ and pseudospin $m_j$ values allowed for one electron. It is
possible and convenient to rewrite this simplified version of the model in
terms of fictitious spin-1 states, the $m_j$ quantum number becoming the
$z$ projection of a pseudospin $S=1$. For this simplified version we can
rewrite the hamiltonian (\ref{gen_model}) replacing the free rotor
operators $L^+,L^-,L^z$ with the generators of the spin 1 algebra
\cite{note1.5} $S^+,S^-,S^z$ respectively
\begin{equation}
	H'_{kin}=-\frac{t}{2} \sum_{<i,j>,\sigma}
c^\dagger_{i ,\sigma} c_{j,\sigma} 		(S^+_i S^-_j + h.c.)
\label{spin1again}
\end{equation}

The full ladder of rotational states (\ref{psih},\ref{psii}) has now
disappeared, being replaced just by the double degeneracy of the $n$=1
sites ($S_z$=$\pm$1), with $H_{rot} \equiv 0$ for all $n$.

The extra terms (\ref{hloc}) would still need to be added to $H'_{kin}$.
However, $H_{rot}$ has the simple effect of giving a energy shift of
$\Omega/4$ per each singly occupied site. As suggested above, this has the
same effect of an operator such as:
\begin{equation}
\Omega \sum_j \left( \frac {n_{j,\uparrow} n_{j,\downarrow}}{2} -
                \frac {n_{j,\uparrow} + n_{j,\downarrow}}{4} \right) \ ,
\end{equation}
i.e. it is the same as a positive Hubbard $U$ term with
$U=\frac{\Omega}{2}$, apart from a chemical potential. For $t<<\Omega$,
this term amounts to a divergent shift of the Hubbard $U$. Such a diverging
term has no physical origin (the JT energy gains we have neglected in
(\ref{enh}) and (\ref{eni}) are also infinite and have opposite sign).
Therefore we will omit it, and simply work with Hamiltonian
(\ref{spin1again}), with the only {\em caveat} that we need to remember the
shift in Hubbard $U$ when comparing the results of this low-cutoff model
(\ref{spin1again}) with the fully converged one.

Both hamiltonians (\ref{gen_model}) and (\ref{spin1again}) show a
significant degree of symmetry which we can take advantage of. Each of them
conserves the number of electrons $N_{el}$, the total pseudo-angular
momentum $ 2 J = M = \sum_i m_i $ (i. e. the total $S_z$ in the $S$=1
pseudospin version), and total electron spin.  For a linear chain, the
lattice translational symmetry is also obvious.  Pseudospin conservation is
a result of our approximation $t_x = t_y$. The others are exact. We choose
to study the problem (\ref{gen_model}) in the manifold at half filling
($N_{el} = N$ ) and at $ M = 0 $ (even $ N$), because of the higher
symmetry present in this case, which includes electron-hole symmetry.
Although we have not yet carried out a complete study of the model away
from half filling, we believe that the basic physics will be (at least for
$U$=0) the same, due to a suggestive analogy with the negative $U$ Hubbard
model which will finally emerge.

At this point, we are set with two alternative working models. The
electron--rotor (ER) model
\begin{equation}
	H_{ER}=H_{kin}+H_{rot}+H_{U} \ ,
\label{HER}
\end{equation}
where $H_{kin}$, $H_{rot}$ and $H_{U}$ are given by (\ref{gen_model}),
(\ref{hloc}), is more realistic, and is characterized by two parameters, the
hopping energy $t$ and the rotor energy $\Omega$. This latter quantity, in
turn, contains the ionic mass, and will therefore make the model sensitive
to isotopic changes. The second, electron--pseudospin one (EP) model
\begin{equation}
        H_{EP}=H'_{kin}+H_{U} \ ,
\label{HEP}
\end{equation}
where $H'_{kin}$ is given by (\ref{spin1again}), represents the extreme
molecular limit and is more idealized, the hopping energy $t$ being the
only parameter. Clearly, there will be no isotope effect in this model.

Although the important terms are $H_{kin}$ and $H'_{kin}$, both models are
endowed with the Hubbard term $H_U$, which can describe additional
repulsive interactions, and is also convenient as a gauge of the effective
attractions which will arise. Having taken $\Omega$ as the energy unit, the
physical results in the ER model will depend on the two dimensionless
ratios $t/\Omega$, $U/\Omega$.  Those in the EP model will depend only upon
$U/t$, making direct comparison with the simple Hubbard model particularly
straightforward.

As it will be shown, there is numerical evidence that the two models, ER
and EP, lead to qualitatively similar effects, at least when $t$ is not too
large. Hence, it will be possible for many purposes to focus on the simpler
EP model.

In the next two sections, we propose to study these models on a 1D linear
chain, as follows. First, we will study numerically some very small
clusters, by direct diagonalization. This will permit a first crude
comparison between ER and EP, and also between them and the simple Hubbard
model. Next, we will introduce a mean-field theory for model EP on the
infinite 1D linear chain. Here, the $S$=1 pseudospin variables can be
approximately integrated out, giving rise to negative effective
electron-electron forward and backward couplings, again suggesting singlet
pairing.

\section{Numerical studies for small linear chain clusters  (rings)}
\label{numeric}

We consider here small $N$-site linear chain clusters (rings), in
particular $N$=4,6,8, accessible to numerical diagonalization using
conventional Lanczos method. While these sizes are admittedly small, we
find that the qualitative results for small $N$ are clear enough at this
initial stage \cite{white}.

We define useful equal--time correlation functions for singlet
superconductivity (SC) and spin density wave (SDW) in the standard form of
$q$-space ''structure factors``:
\begin{eqnarray}
S_{SC}(q) &=&{1 \over N}\sum_{j,l}^N e^{iq(j-l)}
	\left\langle c^\dagger_{j,\downarrow} c^\dagger_{j,\uparrow}
	 	     c_{l,\uparrow} c_{l,\downarrow} ~{\bf I}_m \right\rangle
\label{SC}  \\
S_{SDW}(q) &=&{1 \over N}\sum_{j,l}^N e^{iq(j-l)}
	\left\langle c^\dagger_{j,\uparrow}c_{j,\downarrow}
	     c^\dagger_{l,\downarrow}c_{l,\uparrow} ~{\bf I}_m \right\rangle
\label{SDW}
\end{eqnarray}

We ignore alternative channels, such as charge density waves (CDW) or
triplet superconductivity \cite{solyom}, which can also be probed, but
whose behaviour is not relevant at this stage. In particular, a CDW will be
definitely disfavoured in the more general case away from half filling.  The
property of correlations (\ref{SC}) and (\ref{SDW}) is that they transform
into one another under the transformation $c_{j,\uparrow} \to
\tilde{c}_{j,\uparrow}$, $c^\dagger_{j,\downarrow} \to (-1)^j
\tilde{c}_{j,\downarrow}$, which, remarkably, amounts simply to the
transformation $U\to -U$ \cite{lieb} in the Hubbard model.  Hence in that
model $S_{SC}(q)$ and $S_{SDW}(q)$ are perfectly symmetric around $U$=0,
where their values must cross, the SC instability prevailing for $U<$0, the
SDW for $U>$0. This of course is confirmed for the $N$-site rings, as shown
in Fig. \ref{Hubcorr}. In the upper panel we choose PBC, so that the free
system ($U=0$) is in a closed shell configuration of 6 electrons.  In the
lower panel instead ABC are applied. Here, the resulting shift of the
single-particle $k$-states yields an open shell for free fermions at half
filling. Open shell and closed shell calculations must finally converge to
the same answer for $N\to \infty$, and their systematic comparison at
finite $N$ provides a rough but useful measure of finite-size corrections.

Our strategy is therefore to calculate $S_{SC}$ and $S_{SDW}$ for our
hamiltonian $H_{ER}$ and $H_{EP}$, as a function of $U/t$, and to {\em find
the value of} $U=U^*$ {\em where they cross}, so that superconducting
pairing prevails at all $U<U^*$. The finding that $U^*$ is finally positive
will in turn imply that the bare $U$=0 model is approximately equivalent to
a {\em negative} Hubbard $U$ model, with $U=U_{eff}$, where a crude linear
estimate is
\begin{equation}
	U_{eff} \approx -U^* \ ,
\end{equation}
so long as $U^*$ is small.

Figure \ref{EPcorr} shows results for the EP model, obtained for $N$=4,6,8
sites, at half filling ($N_{el}=N$) as a function of $U$.  The two panels
a) and b) correspond to the different choices of closed shells and open
shells respectively.  For instance, $N=4$ and 8 correspond to closed shells
generated with antiperiodic boundary conditions (ABC), while $N=6$ does
that with periodic boundary conditions (PBC).  Conversely $N=4$ and 8 yield
open shells with PBC, $N=6$ with ABC.  The condensation wavevector $q$ is
correspondingly zero with PBC and $q=\frac{\pi}{N}$ for ABC.

These results show, strikingly, that {\em in the EP model, a finite positive}
$U^*/t$ {\em is needed to suppress superconductivity in favor of SDW's}.
Roughly, the EP model (\ref{spin1again}) behaves therefore like a negative
$U$ Hubbard model, with $U_{eff} \propto - t$ For small $N$, the value of
$U_{eff}$ varies with $N$, and also depends on whether the shell is closed /
open. Although we have not tried a systematic finite-size scaling
extrapolation for $U_{eff}$ to the $N=\infty$ limit, the result up to $N$=8
suggests that
\begin{equation}
	- 0.8 t < U_{eff} < - 0.2 t \ .
\label{Ueff}
\end{equation}
In particular at $N$=6 both ABC and PBC yield $U_{eff}= - 0.37|t|$, which may
therefore be a likely value.

We have also studied the full ER model (\ref{HER}). For this model, the
Hilbert space is that of states (\ref{basis}), with an upper cutoff in the
rotor states $E_{cut}$.  Due to the larger Hilbert space, we
have restricted calculations to $N=4,6$. We proceed by calculating $S_{SC}$
and $S_{SDW}$ for fixed $t/\Omega$ as a function of $U/\Omega$, and we look
for the value $U^*/\Omega$ where they cross. This again defines, via
(\ref{Ueff}), a value for $U_{eff}$. Typical results are shown in Fig.
\ref{ERcorr}.

Now $t$ is an independent parameter. The effective interaction $U_{eff}$
can be recalculated by varying $t$, and the results are given in Fig.
\ref{Uefffig}. The main feature is that the negative $U_{eff}$ at small
$t$, already found in model EP, is confirmed. Hence, model ER also leads to
superconductive pairing for $t/\Omega$ not too large. This fully confirms
our expectations that kinematical restrictions imposed by the switching of
orbital states are important in that regime.  For larger values of $t$,
these restrictions gradually become irrelevant, until, for $t\to \infty$ we
recover the value $U_{eff}\to 0$. In other words, when the hopping energy
is too large, the DJT effect does not work any more, and Fig. \ref{Uefffig}
describes how its ``phase'' part is quenched (the JT distortion magnitude
is held constant in our model).

One may suspect that the pairing we are demonstrating is just a consequence
of the exact symmetry between the $x$ and $y$ degenerate molecular orbitals
that we enforce by the assumption of having equal intermolecular hopping
matrix elements $t_x$, $t_y$, defined in (\ref{hhop}), as discussed in
Section \ref{model}. A simple test releasing this assumption shows that
this is actually not the case. If $t_x \neq t_y$, the hitherto missing term
corresponding to unequal $t_x$, $t_y$ term is the following kinetic
additive contribution:
\begin{equation}
   H_{kin}' =  -{t'\over 2} \sum_{<i,j>,\sigma}
                 c^\dagger_{i,\sigma} c_{j,\sigma}
                ( L^+_i L^+_j + L^-_i L^-_j) \ ,
\label{kinp}
\end{equation}
where $t'$ is the independent hopping amplitude $t'=(t_x-t_y)/2$.  This
term violates the conservation of total $m$ (equivalent to complete $x$-$y$
symmetry).  By adding to the symmetric hamiltonian (\ref{HER}) a term like
(\ref{kinp}) we can break continuously this symmetry, monitoring the effects
of this on pairing, in particular on the $U_{eff}$ that was defined above.

In Fig. \ref{tp_tm} we plot $U_{eff}$ for the EP model (for simplicity) in
the closed shell configurations for 4 and 6 sites, as a function of $t/t'$,
at fixed $t+t'=t_x$.  This figure shows that although the negative
effective attractive $U_{eff}$ is maximum for $t_x=t_y$ and $t_x=-t_y$, it
is not cancelled in the general case $t_x \neq t_y$, except for the very
special case $t_y=0$ (or $t_x=0$). The cancellation of pairing interaction
in this limit is due to the complete breaking of the rotational symmetry,
creating two separate bands from the $x$ and $y$ orbitals, the
$x$-originated being a regular tight-binding band, the $y$ one being made
of localized degenerate states. Anyway, this is indeed a very special case,
in which no $t_y$ term is present. In a more realistic situation having,
say, $t_y\approx - t_x/2$, the pairing effect fully survives.

\section{ The infinite chain -- mean field BCS approach.}
\label{meanf}

In order to get a qualitative idea of the effects of the Berry phase
constraints on the infinite system, we first integrate out the vibron
degrees of freedom in the simple spin--1 model Hamiltonian (\ref{spin1}).
To do this, we make the further approximation of replacing the spin
operators by pseudo--fermion operators representing the vibron excitations.
We can then integrate out the vibron degrees of freedom and apply the BCS
equations to the resulting interacting fermion system.

Introducing an auxiliary spin-$\frac{1}{2}$ fermion $B^\dagger_{i\alpha}$, we
classify the $m=+1$ vibron state as an ``$\alpha$--up" state, and the
$m=-1$ state as an ``$\alpha$--down" state. The $m=0$ state is treated as a
vacuum state for the $B$-fermions, which we will call ``berryons":
\begin{equation}
|m=1>\equiv B^\dagger_\uparrow |0> \ , \ |m=-1>\equiv B^\dagger_\downarrow|0>.
\label{B_repres}
\end{equation}
Using this representation, we express the spin--1 operators by
\begin{equation}
S^+\equiv B^\dagger_\uparrow + B_\downarrow \ , \
S^-\equiv B_\uparrow + B^\dagger_\downarrow \ , \
S^z\equiv B^\dagger_\uparrow B_\uparrow - B^\dagger_\downarrow B_\downarrow \ .
\end{equation}
This representation is overcomplete (in particular it does not exclude
unphysical states with $n_i$, $m_i$ of different parity), and will
therefore not allow a strictly variational treatment. Still, it is of use
in exploring whether the model does or does not display tendencies toward
pairing at the simplest mean field level.

We rewrite the hamiltonian (\ref{spin1}) in this fermion representation as:
\begin{equation}
	H_{kin}=-\frac{t}{2} \sum_{<i,j>,\sigma}
		 c^\dagger_{i ,\sigma} c_{j,\sigma}
		( B^\dagger_{i \uparrow} B_{j\uparrow} +
		  B^\dagger_{i \downarrow} B_{j\downarrow}+
		  B^\dagger_{i \uparrow} B^\dagger_{j\downarrow} +
		  B_{i \downarrow} B_{j\uparrow} + h.c.)
\label{Hdx}
\end{equation}
or, in Fourier representation,
\begin{eqnarray}
	H_{kin}&=&-\frac{t}{2N} \sum_{k_1,k_2,q,\sigma} \cos (k_1+q)
		 c^\dagger_{k_1 ,\sigma} c_{k_2,\sigma} \nonumber\\
	& &	( B^\dagger_{q\uparrow} B_{k_1-k_2+q\uparrow} +
		  B^\dagger_{q\downarrow} B_{k_1-k_2+q\downarrow}+
		  B^\dagger_{q\uparrow} B^\dagger_{-k_1+k_2-q\downarrow} +
		  B_{-q\downarrow} B_{k_1-k_2+q\uparrow} + h.c.) \ .
\label{Hdk}
\end{eqnarray}

We take as a zeroth order  mean field hamiltonian just free fermions (note that
this term was missing in the original problem)
\begin{equation}
	H_{MF}= \sum_{k,\sigma} \epsilon_k
		 c^\dagger_{k,\sigma} c_{k,\sigma} +
	   \sum_{k,\alpha} \eta_k
		 B^\dagger_{k,\alpha} B_{k,\alpha}
\label{Hmf}
\end{equation}
such that
\begin{equation}
	E_{MF}= < 0_{MF}| H | 0_{MF}>
\label{Emf1}
\end{equation}
is minimum. $| 0_{MF}>$ is the direct product of a half filled Fermi sea of
$c$-electrons and of a Fermi sea of berryons filled up to $x\equiv
N_B/N$.  Here, $x$ can be regarded as an adjustable variational parameter
(although $H_{MF}$ is not truly variational, as it violates the constraints
required by Eq. (\ref{const})).
The precise value of $x$ is however immaterial,
since the qualitative results we will find appear to be independent of $N_B$.

Direct substitution gives
\begin{equation}
	E_{MF}= -\frac{8tN}{\pi^2} \sin (\frac{\pi}{2} x) \ ,
\label{Emf2}
\end{equation}
which is minimum for $x=1$. The single particle excitation energies are
\begin{eqnarray}
	\epsilon_k&=&-2\theta \cos k \\
	\eta_k	&=&-\frac{4t}{\pi}\cos k \ ,
\label{singlep}
\end{eqnarray}
where $\theta\equiv \frac{2t}{\pi}\sin (\pi x /2)$ is the effective mean
field hopping  amplitude for $c$-electrons.

The next step is the determination of the first nontrivial correction to
the mean field due to the actual interaction (\ref{Hdk})
between $c$-electrons and
berryons. These corrections are achieved through an expansion in the
interaction $(H-H_{MF})$ around the free dynamics $H_{MF}$.
 For this purpose, we write
the full many-body partition function at temperature $1/\beta$
\cite{negele}:
\begin{equation}
	Z=Z_{MF} \left< \left< e^{-S}
		\right>_B \right>_c
	\approx Z_{MF} \left< e^{-<S>_B+\frac{1}{2}[<S^2>_B-<S>_B^2]+...}
			  \right>_c \ ,
\label{Z}
\end{equation}
with
\begin{eqnarray}
	Z_{MF}&=&Z_{MF}^c Z_{MF}^B		\nonumber\\
	S&=&\int_0^\beta d\tau
	 [H(c(\tau),...B(\tau))-H_{MF}(c(\tau),...B(\tau))] \nonumber\\
	<O[c_{k\sigma}]>_c&=&
 \int \frac{ D[c^\dagger_{k\sigma} c_{k\sigma}]} {Z^c_{MF}}
 e^{-\int_0^\beta d\tau \sum_{k,\sigma} c^\dagger_{k\sigma}(\tau)
 (\partial_\tau + \epsilon_k)c_{k\sigma}(\tau)}O[c_{k\sigma}(\tau)]\nonumber\\
	<O[B_{k\alpha}]>_B & = &
 \int \frac{ D[B^\dagger_{k\alpha} B_{k\alpha}]}{Z^B_{MF}}
 e^{-\int_0^\beta d\tau \sum_{k,\alpha} B^\dagger_{k\alpha}(\tau)
 (\partial_\tau + \eta_k)B_{k\alpha}(\tau)} O[B_{k\alpha}(\tau)]
\end{eqnarray}
where $O[.]$ is any operator, and
\begin{equation}
	Z_{MF}^c=
 \int  D[c^\dagger_{k\sigma} c_{k\sigma}]
 e^{-\int_0^\beta d\tau \sum_{k,\sigma} c^\dagger_{k\sigma}(\tau)
 (\partial_\tau + \epsilon_k)c_{k\sigma}(\tau)}
\end{equation}
and a similar expression for $Z_{MF}^B$.

Averaging over the non-interacting many body $B$-fields in the cumulant
expansion in (\ref{Z}) leaves an effective hamiltonian operator for the
$c$-electrons. That expansion contains a first term $<S>_B$ whose form is
$\int_0^\beta d\tau \sum c^\dagger_{k\sigma} (\tau) c_{k\sigma} (\tau)$.
It simply renormalizes the mean field parameters.  The lowest order
nontrivial action correction belongs to $-<S^2>_B$, having the form of an
effective electron-electron interaction term
\begin{equation}
S_{eff}=\frac{1}{N} \int d\tau \int d\tau' \sum_{\sigma,\sigma'}
\sum_{k_1,k_2,k_3,k_4} c^\dagger_{k_1\sigma}(\tau) c_{k_2\sigma}(\tau)
c^\dagger_{k_3\sigma'}(\tau') c_{k_4\sigma'}(\tau')
K_{k_1,k_2,k_3,k_4}(\tau - \tau') \ ,
\label{seff}
\end{equation}

This term has a very simple significance. It corresponds to the exchange of
a berryon particle-hole pair with singlet total pseudospin between the two
electrons, as in the diagram of Fig. \ref{diagram}.

The imaginary time integration can be recast in a Matsubara frequency
summation, in terms of a kernel
\begin{eqnarray}
	K_{k_1,k_2,k_3,k_4}(i\omega_B)&=&
	\delta_{k_1+k_3,k_2+k_4} \frac{t^2}{2N}
	\sum_k \cos(k_1+k) \cos(k_3+k+q)
						 \nonumber\\ & &
	\frac{1}{\beta} \sum_{\omega_n}
	[-2\tilde{g}_{k\uparrow}(\omega_n)
		\tilde{g}_{k\uparrow}(\omega_n+\omega_B) +
						 \nonumber\\ & &
		\tilde{g}_{k\uparrow}(\omega_n)
                \tilde{g}_{k\downarrow}(-\omega_n-\omega_B)
		\tilde{g}_{k\downarrow}(\omega_n)
                \tilde{g}_{k\uparrow}(-\omega_n+\omega_B) ] \ ,
\label{kappa1}
\end{eqnarray}
where $q:=k_2-k_1$, $\omega_n$ are fermionic Matsubara frequencies, and
$\tilde{g}_{k\alpha}(\omega_n)$ is the free fermion propagator in Matsubara
space as defined in \cite{negele}.  The sum over the Matsubara frequencies
can be performed to recast Eq.  (\ref{kappa1}) in form
\begin{eqnarray}
	K_{k_1,k_2,k_3,k_4}(i\omega_B)&=&
	\delta_{k_1+k_3,k_2+k_4} \frac{t^2}{2N}
	\sum_k \cos(k_1+k) \cos(k_3+k+q) \nonumber\\ & &
	\left[	2\Sigma(i\omega_B;\xi_k,\xi_{k-q})
		-\Sigma(i\omega_B;\xi_k,-\xi_{k-q})
		-\Sigma(i\omega_B;-\xi_k,\xi_{k-q}) \right] \ ,
\label{kappa2}
\end{eqnarray}
where $\Sigma(z,a,b):=[f_F(a)-f_F(b)]/[z-(b-a)]$, $f_F()$ are Fermi
occupation factors and $\xi_k:=\eta_k-\mu=
-\frac{4t}{\pi}[\cos(k)-\cos(\frac{\pi x} {2})]$ are the single--particle
excitation energies for the berryons reduced by the corresponding
chemical potential.

We would now like to extract physical conclusions from this calculation.
Since we deal with an effective 1D electron system, we wish to use
the calculated effective electron-electron scattering as a guide to
understanding which one of the standard 1D Luttinger model fixed points
will prevail.

In particular, for that model, an estimate of the forward and backward
coupling constants $g_1$ and $g_2$ \cite{solyom} will determine what kind
of ground state to expect.

The pair scattering amplitude we have obtained is obviously time- (or
frequency-) dependent, i.e., non-hamiltonian in nature. In this sense,
straight identification with true hamiltonian parameters such as $g_1$ and
$g_2$ \cite{solyom} is not automatically correct.  However, we see no
physical reason preventing us from using our derived amplitudes as
effective coupling constants so long as we stay sufficiently close to the
Fermi surface.

We therefore identify
\begin{equation}
	g_1\propto K_{-k_F,k_F,k_F,-k_F}(i\omega_B) \ \ (q=\pi) \ , \
	g_2\propto K_{k_F,k_F,-k_F,-k_F}(i\omega_B) \ \ (q=0)   \ .
\label{g1g2}
\end{equation}

Direct computation of $K$ for these special momenta gives:
\begin{eqnarray}
K_{-k_F,k_F,k_F,-k_F}(z)&=&-\frac{t^2}{2N}\int_0^\pi \frac{dk}{\pi} \sin^2(k)
		[ f_F(-\xi_k) - f_F(\xi_k) ] \frac{2}{2\xi_k - z} \nonumber\\
K_{k_F,k_F,-k_F,-k_F}(z)&=&-\frac{t^2}{2N}\int_0^\pi \frac{dk}{\pi} \sin^2(k)
                [ f_F(-\xi_k) - f_F(\xi_k) ]\frac{4\xi_k}{4\xi_k^2 -z^2}
\label{2kappi}
\end{eqnarray}

Interestingly, these two couplings arise from different terms in the
hamiltonian.The former is due to the $B^\dagger B$, $B ~ B^\dagger$ terms
in (\ref{Hdk}), whereas the latter to the $B^\dagger B^\dagger$ and $B~B$
terms.

In the zero frequency limit $z\to 0$, the two quantities are negative and
coincide. In terms of the Luttinger model phase diagram \cite{solyom}, this
corresponds to a spin singlet superconducting state , which is therefore
found to prevail.  This is in very good agreement with the effective
negative $U$ of the previous section, with the additional remark that the
alternative possibility of charge-density waves is now explicitly ruled
out.

 Actually, the zero frequency limit is singular at zero temperature, when
the Fermi functions become step functions and the $k$ integration diverges
logarithmically for vanishing frequency around $k_F^B$, the Fermi momentum
for the berryons. In other words, in this approximation $g_1=g_2$
diverge as $\ln |z|$, at small $z$.  This is a singular feature, due to our
assuming the exchange of a bare, unrenormalized particle-hole pair as in
Fig. \ref{diagram}. Higher order diagrams will modify that.  More
importantly, in presence of a finite pairing amplitude, for example, this
divergence will disappear, due to a pairing gap in the berryon
spectrum. There is in fact an exact symmetry between fermions and
pseudofermions, and the two Cooper channels are also identical.

\section{Discussion}

A very important property of our Berry phase--constrained tunnelling
Hamiltonian is the fact that the constraint operates at the energy scale of
the tunnelling matrix element $t$. Thus the pairing tendency induced by the
constraints is not dictated directly by the strength of the
intra--molecular electron--vibron coupling, but rather by the indirect
effect of this coupling in the semi--classical limit on the relative phase
space available for single electron tunnelling versus that for pair
tunnelling. Thus the enforcing of the Berry phase constraint effectively
separates the energy scale of the tunnelling, $t$, from that of the
internal degrees of freedom of the constituent molecules.

In the case of C$_{60}$ itself, the model is in too extreme a
semi--classical limit to give a reasonable representation of the physics of
K$_3$C$_{60}$ since, in that case, the dimensionless electron--vibron
coupling strength $g$ is of order $\sim 0.4$, whereas the strong coupling
limit where the Berry phase representation becomes useful is for $g
\gtrsim 1$.

Nevertheless, the model does illustrate a new physical principle for
superconductivity in strongly constrained systems. It is tempting to make
an analogy with the physics of the $t-J$ model of interest for describing
the physics of doped Mott insulators. In that case the hopping Hamiltonian
my be re-written as
\begin{equation}
H_t = -t\sum_{<ij>}(S^+_i c^\dagger_{i\downarrow} c_{j\downarrow}S^-_j
 + S^-_i c^\dagger_{i\uparrow} c_{j\uparrow}S^+_j + h.c.)
\end{equation}
 where $S^+_i, S^-_j$ are spin raising and lowering operators. A number of
recent studies \cite{t-J} suggest that pairing of holes close to the
half-full insulating state occurs as a result of kinematical constraints in
this model.

Although the present model does not have the ordered or quasi--ordered
background of the antiferromagnetic state in the $t-J$ model, it does have
the feature that pair tunnelling proceeds by each partner causing a vibron
excitation when executing a tunnelling step, which then annihilate when the
pair of carriers come together again on the same site. Similarly, in the
$t-J$ model case, individual hole hopping is accompanied by a spin--flip,
which can the be cancelled by the hopping of its partner.

It is of interest to consider whether our Berry phase considerations could
also apply to the Chevrel--phase class of superconductors such as
LaMo$_6$Se$_8$ or PbMo$_6$S$_y$Se$_{8-y}$ \cite{chevrel}.  In these
materials the Mo$_6$Se$_8$ cluster has a set of degenerate LUMO orbitals
analogous to those in C$_{60}$. Measurements of the doping dependence of
$T_c$ indicate a sharp maximum as a function of doping in the unfilled LUMO
shell. Thus there is the possibility of a general class of
constraint--driven superconductors with distinctly different dependence of
$T_c$ on material parameters than those of the conventional BCS--type
electron--phonon superconductors \cite{uemura}.

Although, as stated above, we do not expect our model to be a realistic
representation of the physics of K$_3$C$_{60}$, it would nevertheless be
interesting to test experimentally whether the kind of electron--vibron
coupling we have proposed could be observed in this compound. One way to do
this would be through the two--vibron Raman spectrum.  Our coupling
mechanism would naturally lead to a direct electron--hole pair channel
coupling to a pair of vibron modes. This channel would open up a gap of
$2\Delta$, in the vibron spectrum where $\Delta$ is the superconducting
gap, on lowering the temperature of the material below $T_c$.

More generally, we observe that the pseudorotor Berry phase mechanism
sketched here, ties together electron hopping with the hopping of quanta of
{\em orbital} molecular angular momentum, which is unquenched in the
free-molecule limit we start from. In the paired state orbital quanta are
also paired, whereby the {\em orbital excitation branch will also develop a
gap at} $q= 2k_F$. The gap will follow identically the superconducting gap
at $T= T_c$.  The staggered orbital susceptibility should therefore be
maximum at $ T_c$.  In turn, the uniform $q=0$ orbital susceptibility may
also develop a maximum, although weaker, via momentum nonconserving or
local field effects.  It is possible that orbital effects of this kind,
even if weaker than suggested by this extreme picture, could be detectable,
e.g., by NMR. In particular, the relaxation time $1/T_1$ could be enhanced
at low temperature, and peak up around $T_c$ due to large susceptibility
fluctuations \cite{rigamonti}.

It would be of considerable interest to see if these new effect could be
observed in K$_3$C$_{60}$, and Rb$_3$C$_{60}$.  Encouragingly, in this
latter compound, very recent NMR data \cite{zimmer} seem to indicate a
behavior of the relaxation time which is anomalous precisely in the way
suggested above. The anomaly at $T_c$, in particular, is seen in the Rb
ion, but not on the carbon, as we would expect for a C$_{60}$ orbital
effect.

Finally, it is of interest to speculate that $T_c$ would be enhanced by
doping our model system away from half--filling. Because of the nature
of our pairing mechanism, the carriers would have more phase space for
pairing if each partner in a pair could find many empty neighboring sites
to hop on to before re--pairing. Thus the doping dependence of $T_c$ might
be expected to have a maximum away from half--filling in systems
for which this mechanism is driving the superconductivity. In this sense
the case of half filling is probably the least favorable. There are
indications from an exact solution for the 2--electron state that in model
EP, $U_{eff}$ is one order of magnitude more attractive near zero filling
\cite{santoro}.

In the fullerides, exact half filling appears to be required by chemical
stability \cite{note2}. In the Chevrel systems, however, where continuous
doping is feasible, one indeed finds a maximum of $T_c$ for a hole density
close to one per molecular unit \cite{chevrel}. This corresponds to only
1/6 filling of the narrow $\Gamma_{25}$ molecular band in that case
\cite{mattheiss}. A second observation is that the correlation length
should tend to be short, of the order of the intermolecular distance $a$,
since this is the scale where the energy gain takes place.  In
K$_3$C$_{60}$, this expectation is well borne out, with a correlation
length of order $\equiv 2a$.

Experimentally, it would be of interest to consider building new molecular
solids where high-symmetry Jahn-Teller molecules can exchange electrons.
Larger molecules may be better ones because of a weaker intramolecular
Coulomb repulsion $U$. Relatively weak JT coupling may provide an
additional favourable circumstance, since in that case the effective
$\Omega$ is larger (although our treatment does not strictly apply there)
and DJT quantum effects are more important. Both these conditions are met
in the fullerides, but it might be possible to find other systems where
they apply.

\section{Conclusions}

We have proposed a model for constrained tunnelling of charge carriers in
a lattice of Berry phase molecules, inspired by the physics of the fullerides.

The general model (Eq. \ref{gen_model}) is based on an electron--quantum
rotor Hamiltonian which includes a (in principle infinite) manifold of
vibron states on each site. Although we have been able to investigate the
effects of a large number of vibron states on the pairing tendency, it is
clear that the effect is strongest when only the lowest are important.  For
the extreme case where only the lowest vibron state is important ($S=1$
pseudospin model), both our numerical studies on small clusters (Section
\ref{numeric}) and our BCS--type mean field treatment (Section \ref{meanf})
indicate a strong pairing tendency for a half--filled band. The
fact that the model exhibits intrinsic pairing even in the presence of
Hubbard repulsion $U$ of order the tunnelling matrix element $t$ is
understood readily from the form of the model Hamiltonian (Eq. \ref{spin1},
\ref{gen_model}).  On integrating out the vibron degrees of freedom, we
obtain an effective BCS attraction of order $t^2/W$, where $W$ is a vibron
bandwidth of order $t$.  Thus the Berry phase constraint leads to an
electron--electron attraction whose energy scale is not directly related to
the strength of the intra--molecular electron--vibron coupling in the
semi--classical limit.

In physical terms, our model is based on the entanglement of orbital
angular momentum of the individual molecules with electron hopping between
molecules. Pairing of electrons is generated by an accompanying
(``singlet'') pairing of orbital momenta on neighbouring
molecules, suggesting short correlation lengths in the order of the
intermolecular spacing. It has been argued that this mechanism might be
relevant also in such other molecular superconductors, such as the Chevrel
compounds.

\section{Acknowledgments}

It is a special pleasure to acknowledge many illuminating discussion with
and much help from G. Santoro and A. Parola, and also from M. Airoldi and
M. Fabrizio. The sponsorship of NATO, through CRG 920828 and of EEC through
Contract ERBCHRXCT420062, are gratefully acknowledged.

\begin{figure}
\caption{Structure factors $S_{SC}(q)$
(\protect\ref{SC}) and $S_{SDW}(q)$ (\protect\ref{SDW}) for the Hubbard
model (6 sites ring), as a function of the dimensionless parameter $U/t$.
In the upper panel periodic boundary conditions (PBC) are applied to
indexes, so that for $U$=0 the ground state is nondegenerate (``closed
shell''), while in the lower panel antiperiodic boundary conditions (ABC)
make the noninteracting ground state degenerate (``open shell''). In panel
a) $q=0$, while in panel b) $q=\pi/N$, as appropriate to the boundary
conditions applied. Solid dots mark crossings, where the switching from
superconductivity to spin-density waves takes place.}
\label{Hubcorr}
\end{figure}

\begin{figure}
\caption{Structure factors $S_{SC}$ (\protect\ref{SC}) and
$S_{SDW}$ (\protect\ref{SDW}) for the EP model, plotted as a function of
$U/t$. In panel a) we report the result for the closed shell case, while
panel b) has the result for the open shell case, as discussed in the text.
Solid dots mark crossings defining $U^*$, indicating that superconductive
pairing prevails even at positive $U$.}
\label{EPcorr}
\end{figure}

\begin{figure}
\caption{Structure factors $S_{SC}$ (\protect\ref{SC}) and
$S_{SDW}$ (\protect\ref{SDW}) for the ER model ($N$=4,6), plotted as a
function of $U/\Omega$, with $t=1$. In panel a) we report the result for
the closed shell case, while panel b) has the result for the open shell
case. $E_{cut}=6$ for $N$=4, and $E_{cut}=3$ for $N$=6. Solid dots mark
crossings defining $U^*$, again indicating pairing even at positive $U$.}
\label{ERcorr}
\end{figure}

\begin{figure}
\caption{The effective Hubbard term $U_{eff}$
(\protect\ref{Ueff}) for the ER model ($N$=4,6), plotted as a function of
$t/\Omega$. The $N$=4 ABC and $N$=6 PBC are closed shells, while $N$=4 PBC
and $N$=6 ABC are open shell cases.  $E_{cut}=8$ for $N$=4, and $E_{cut}=4$
for $N$=6. For $t/\Omega<<1$, the slope of $U_{eff}$ coincides with
$U_{eff}/t$ of the EP model (Fig. \protect\ref{EPcorr}). For $t/\Omega>>1$,
the dynamical Jahn-Teller effects are suppressed, whence $U_{eff}\to 0$.}
\label{Uefffig}
\end{figure}

\begin{figure}
\caption{The effective Hubbard term $U_{eff}$
(\protect\ref{Ueff}) for the EP model ($N$=4,6, in closed shells) with the
addition of the term (\protect\ref{kinp}), describing $t_x \neq t_y$,
plotted as a function of $t/t'$, at fixed $t+t'=t_x=1$. The effective
interaction survives everywhere except at the isolated point $t/t'=1$,
corresponding to the unphysical case $t_y=0$ (or $t_x=0$).}
\label{tp_tm}
\end{figure}

\begin{figure}
\caption{The second order effective interaction between electrons is due to
the exchange of a pair of berryons.}
\label{diagram}
\end{figure}


\begin{references}
\bibitem{note1} For a more general lattice symmetry, there will be also terms
$ c^\dagger_x c_y $ but we stick for convenience to the simpler high
symmetry case for the time being.
\bibitem{note1.5} Although $L^+$ and $L^-$ commute in the full rotor space,
it is easy to verify that $[S^+,S^-]=S^z$, as appropriate in a spin
algebra.
\bibitem{white} For a better understanding of the large $N$ thermodynamic
limit, we are presently considering finite-size scaling studies of this model.
\bibitem{note2} The insulating compounds K$_4$C$_{60}$,
and Rb$_4$C$_{60}$, on the other hand, cannot really be thought as the
result of doping K$_3$C$_{60}$, and Rb$_3$C$_{60}$ away from half filling.
We have in fact suggested that their insulating behavior should be of
Jahn-Teller nature \cite{assa2}.
\end{references}
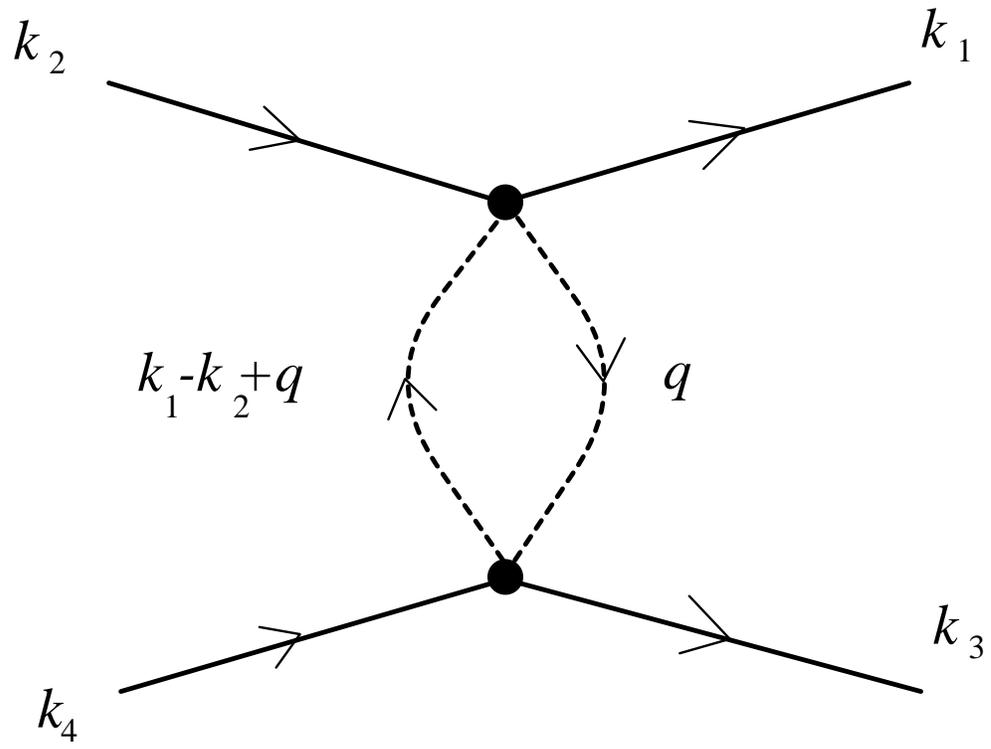
\end{document}